\documentclass[sigconf]{acmart}
\usepackage{bm}
\usepackage{multirow}
\usepackage{subfigure}
\usepackage{algorithmic}
\usepackage[ruled, linesnumbered]{algorithm2e}
\usepackage{booktabs} 
\usepackage{amsmath}

\usepackage{amssymb}
\usepackage{xcolor}
\usepackage{multirow}
\usepackage{graphicx}
\usepackage{subfigure}
\usepackage{amsfonts}
\usepackage{balance}

\newcommand{\fullname}{\emph{Variational Graph Generative-Contrastive Learning~(VGCL)}}
\newcommand{\shortname}{\emph{VGCL}}

\AtBeginDocument{%
  \providecommand\BibTeX{{%
    \normalfont B\kern-0.5em{\scshape i\kern-0.25em b}\kern-0.8em\TeX}}}

\copyrightyear{2023}
\acmYear{2023}
\setcopyright{acmlicensed}\acmConference[SIGIR '23]{Proceedings of the 46th International ACM SIGIR Conference on Research and Development in Information Retrieval}{July 23--27, 2023}{Taipei, Taiwan}
\acmBooktitle{Proceedings of the 46th International ACM SIGIR Conference on Research and Development in Information Retrieval (SIGIR '23), July 23--27, 2023, Taipei, Taiwan}
\acmPrice{15.00}
\acmDOI{10.1145/3539618.3591691}
\acmISBN{978-1-4503-9408-6/23/07}
\begin{document}

\title{Generative-Contrastive Graph Learning for Recommendation}

\author{Yonghui Yang}
\affiliation{
\department{Key Laboratory of Knowledge Engineering with Big Data,}
\institution{Hefei University of Technology}
\country{}
}
\email{yyh.hfut@gmail.com}

\author{Zhengwei Wu}
\affiliation{
\institution{Ant Group}
\country{}
}
\email{zejun.wzw@antfin.com}

\author{Le Wu}
\authornotemark[1]
\affiliation{
\department{Key Laboratory of Knowledge Engineering with Big Data,}
\institution{Hefei University of Technology}
\country{}
}
\email{lewu.ustc@gmail.com}
\thanks{This work is done when Yonghui Yang works as an intern at Ant Group. \\ Le Wu is the Corresponding author}

\author{Kun Zhang}
\affiliation{
\department{Key Laboratory of Knowledge Engineering with Big Data,}
\institution{Hefei University of Technology}
\country{}
}
\email{zhang1028kun@gmail.com}

\author{Richang Hong}
\affiliation{
\department{Key Laboratory of Knowledge Engineering with Big Data,}
\institution{Hefei University of Technology}
\country{}
}
\email{hongrc.hfut@gmail.com}

\author{Zhiqiang Zhang}
\affiliation{
\institution{Ant Group}
\country{}
}
\email{lingyao.zzq@antfin.com}

\author{Jun Zhou}
\affiliation{
\institution{Ant Group}
\country{}
}
\email{jun.zhoujun@antfin.com}

\author{Meng Wang}
\affiliation{
\department{Key Laboratory of Knowledge Engineering with Big Data,}
\institution{Hefei University of Technology}
\institution{Institute of Artificial Intelligence, Hefei Comprehensive National Science Center}
\country{}
}
\email{eric.mengwang@gmail.com}

\renewcommand{\shortauthors}{Yonghui Yang et al.}


\begin{abstract}

By treating users' interactions as a user-item graph, graph learning models have been widely deployed in Collaborative Filtering~(CF) based recommendation.  Recently, researchers have introduced Graph Contrastive Learning~(GCL) techniques into CF to alleviate the sparse supervision issue, which first constructs contrastive views by data augmentations and then provides self-supervised signals by maximizing the mutual information between contrastive views.
Despite the effectiveness, we argue that current GCL-based recommendation models are still limited as current data augmentation techniques, either structure augmentation or feature augmentation. First, structure augmentation randomly dropout nodes or edges, which is easy to destroy the intrinsic nature of the user-item graph. Second, feature augmentation imposes the same scale noise augmentation on each node, which neglects the unique characteristics of nodes on the graph. 

To tackle the above limitations, we propose a novel \fullname~ framework for recommendation. Specifically, we leverage variational graph reconstruction to estimate a Gaussian distribution of each node, then generate multiple contrastive views through multiple samplings from the estimated distributions, which builds a bridge between generative and contrastive learning. The generated contrastive views can well reconstruct the input graph without information distortion. Besides, the estimated variances are tailored to each node, which regulates the scale of contrastive loss for each node on optimization. Considering the similarity of the estimated distributions, we propose a cluster-aware twofold contrastive learning, a node-level to encourage consistency of a node's contrastive views and a cluster-level to encourage consistency of nodes in a cluster. Finally, extensive experimental results on three public datasets clearly demonstrate the effectiveness of the proposed model.

\end{abstract}


\maketitle

\section{Introduction}
CF-based recommendation relies on the observed user-item interactions to learn user and item embeddings for personalized preference prediction, and has been pervasive in real-world applications~\cite{su2009survey}. Early works leverage the matrix factorization technique to obtain user and item embeddings, and then compute users' preferences by inner product~\cite{NIPS2008PMF, UAI2009BPR} or neural networks~\cite{WWW2017NCF}. As users' interactions can be naturally formulated as a user-item graph, borrowing the success of Graph Neural Networks~(GNNs), graph-based CF models have been widely studied with superior performances~\cite{SIGIR2019NGCF, AAAI2020LRGCCF, LightGCN}. These models iteratively propagate the neighborhood information for embedding updates, such that the higher-order collaborative signals can be incorporated for better user and item embedding learning.


Despite the effectiveness, graph-based CF models suffer from the sparse supervision issue for model learning.  As an alternative, self-supervised learning leverages the input data itself as the supervision signal and has attracted many researchers~\cite{jaiswal2020survey, wu2021self}. Among all self-supervised learning, contrastive learning is a popular paradigm that constructs data augmentations to teach the model to compare similar data pairs, and has shown competitive performance in computer vision, natural language processing, graph mining, and so on~\cite{jaiswal2020survey,liu2021self}. Some recent studies have introduced contrastive learning in graph-based CF ~\cite{sgl, ncl, simgcl}. In addition to the supervised recommendation task, GCL-based models first construct multiple contrastive views through data augmentation, and then maximize the mutual information to encourage the consistency of different views.  Existing GCL-based CF methods can be classified into two categories: structure augmentation and feature augmentation. Specifically, structure augmentation randomly dropout graph nodes or edges to obtain subgraph structure, and then feeds the augmented graphs into an encoder for contrastive representations~\cite{sgl}. Feature augmentation adds random noises to node embeddings as contrastive views~\cite{simgcl}. These GCL-based CF models learn self-supervised signals based on data augmentation and significantly improve recommendation performances.


Although data augmentation is the key to the performance of GCL-based CF models, we argue that current solutions are still limited by current data augmentation strategies, either structure augmentation or feature augmentation.
Firstly, structure augmentation randomly dropout nodes or edges, which is easy to destroy the intrinsic nature of the input graph. The reason is that all nodes are connected on the graph and don't satisfy the IID assumption.
Secondly, feature augmentation adds the same scale noise to each node, which neglects the unique characteristics of nodes on the graph.
In real-world recommender systems, different users~(items) have different characteristics, and the data augmentation techniques should be tailored to each user. E.g., some users have more item links in the user-item graph, which contains more supervised signals compared to users with only very few links. How to better exploit the user-item graph structure to design more sophisticated contrastive view construction techniques is still open. 

In this paper, we exploit the potential of the generative model to facilitate contrastive view generation without data augmentation. Specifically, we propose a \fullname~ framework for recommendation. Instead of data augmentation, we leverage variational graph inference~\cite{kipf2016vgae} to estimate a Gaussian distribution of each node, then generate multiple contrastive views through multiple samplings from the estimated distributions. As such, we build a bridge between the generative and contrastive learning models for recommendation. The generated contrastive views can well reconstruct the input graph without information distortion. Besides, the estimated variances are tailored to each node, which can adaptively regulate the scale of contrastive loss of each node for optimization. We consider that similar nodes are closer in the representation space, and then propose cluster-aware contrastive learning with  twofold contrastive objectives. The first one is a node-level contrastive loss that encourages the consistency of each node's multiple views. The second one is a cluster-level contrastive loss that encourages the consistency of different nodes in a cluster, with the cluster learned from the estimated distributions of nodes. The major contributions of this paper are summarized as follows:

 
\begin{itemize}
    \item We introduce a novel generative-contrastive graph learning paradigm from the perspective of better contrastive view construction, and propose a novel \fullname~ framework for recommendation.
    \item We leverage variational graph reconstruction to generate contrastive views, and a design cluster-aware twofold contrastive learning module, such that the self-supervised signals can be better mined at different scales for GCL-based recommendation.
    \item Extensive experiments on three public datasets clearly show the effectiveness of the proposed framework, our \shortname~ consistently outperforms all baselines. 
\end{itemize}

\section{Preliminaries}
\subsection{Graph based Collaborative Filtering}
In fundamental collaborative filtering, there are two kinds of entities: a userset $U$~($|U|=M$) and an itemset $V$~($|V|=N$). Considering the recommendation scenarios with implicit feedback, we use matrix {\small{$\mathbf{R} \in \mathbb{R}^{M\times N}$}} to describe user-item interactions, where each element $\mathbf{r}_{ai}=1$ if user $a$ interacted with item $i$, otherwise $\mathbf{r}_{ai}=0$. Graph-based CF methods~\cite{SIGIR2019NGCF, AAAI2020LRGCCF, LightGCN} formulate the available data as a user-item bipartite graph $\mathcal{G}=\{U \cup V, \mathbf{A}\}$, where $U \cup V$ denotes the set of nodes, and $\mathbf{A}$ is the adjacent matrix defined as follows:

\begin{small}
\begin{flalign}\label{eq:adj_matrix}
\mathbf{A}=\left[\begin{array}{cc}
\mathbf{0}^{M\times M} & \mathbf{R}\\
\mathbf{R}^T & \mathbf{0}^{N\times N}
\end{array}\right].
\end{flalign}
\end{small}

\noindent Given the initialized node embeddings $\mathbf{E}^0$, graph-based CF methods update node embeddings through multiple graph convolutions:
\begin{small}
\begin{flalign}\label{eq:aggregation}
\mathbf{E}^{l}=\mathbf{D}^{-\frac{1}{2}}\mathbf{A}\mathbf{D}^{-\frac{1}{2}} \mathbf{E}^{l-1}, 
\end{flalign}
\end{small}

\noindent where $\mathbf{D}$ is the degree matrix of graph $\mathcal{G}$, $\mathbf{E}^{l}$ and $\mathbf{E}^{l-1}$ denote node embeddings in $l^{th}$ and ${(l-1)}^{th}$ graph convolution layer, respectively. When stacking $L$ graph convolution layers, the final node representations can be obtained with a readout operation:
\begin{small}
\begin{flalign}\label{eq:readout}
\mathbf{E}=Readout(\mathbf{E}^0, \mathbf{E}^1, ..., \mathbf{E}^L).
\end{flalign}
\end{small}

\noindent The pairwise ranking~\cite{UAI2009BPR} loss is adopted to optimize model parameters:
\begin{small}
\begin{equation}
    \label{eq: loss_rec}
	\mathcal{L}_{rec}=\sum_{a=0}^{M-1}\sum\limits_{(i,j)\in D_a }-log\sigma(\hat{r}_{ai}-\hat{r}_{aj}) + \lambda ||\mathbf{E}^0||^2,
\end{equation} 
\end{small}

\noindent where $\sigma(\cdot)$ is the sigmoid activation function, $\lambda$ is the regularization coefficient.
{\small$D_a=\{(i,j)|i\in R_a\!\wedge\!j\not\in R_a\}$} denotes the pairwise training data for user $a$. {\small$R_a$} represents the item set that user $a$ has interacted.


\begin{small}
\begin{figure} [t]
  \begin{center}
    \includegraphics[width=85mm]{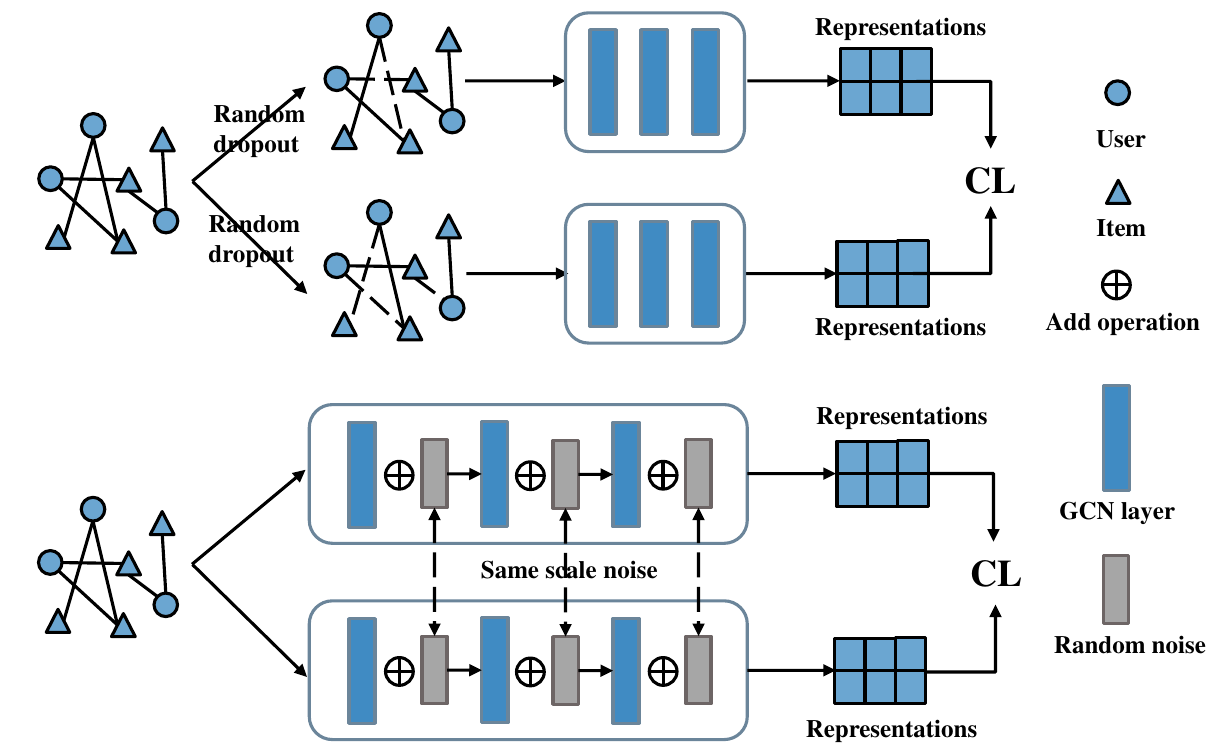}
  \end{center}
    \vspace{-0.2cm}
  \caption{\small{Graph contrastive learning paradigms with structure and feature data augmentation.}}\label{fig: intro}
    \vspace{-0.3cm}
\end{figure}
\end{small}

\subsection{Graph Contrastive Learning for Recommendation}
GCL usually as an auxiliary task to complement recommendation with self-supervised signals~\cite{sgl,ncl,simgcl}, which performs multi-task learning:
\begin{small}
\begin{flalign} \label{eq: multi-task}
    \mathcal{L}=\mathcal{L}_{rec} + \alpha \mathcal{L}_{cl},    
\end{flalign}
\end{small}

\noindent  where $\alpha$ is a hyper-parameter that controls the contrastive task weight, $\mathcal{L}_{cl}$ is the typical InfoNCE loss function~\cite{gutmann2010noise}:
\begin{small}
\begin{flalign} \label{eq: infonce}
    \mathcal{L}_{cl} = \sum\limits_{i \in \mathcal{B}}
    -log \frac{exp({\mathbf{e}'_i}^T\mathbf{e}''_i/\tau)}{\sum\limits_{j \in \mathcal{B}}
    exp({\mathbf{e}'_i}^T\mathbf{e}''_j/\tau)},
\end{flalign}
\end{small}

\noindent where $\mathcal{B}$ denote a batch users~(items), $\tau$ is the contrastive temperature. For node $i$, $e'_i$ and $e''_i$ denote the corresponding contrastive representations with $L_2$ normalization, the same as node $j$. This objective encourages consistency of contrastive representations for each node.

Revisiting GCL-based recommendation models from a data augmentation perspective, there are two popular strategies: structure augmentation~\cite{sgl} and feature augmentation~\cite{simgcl}. As illustrated in the upper part of Figure\ref{fig: intro}, structure augmentation randomly perturb graph structure to obtain two augmented views $\mathcal{G}', \mathcal{G}''$, then generate contrastive representations as follows: 
\begin{small}
\begin{flalign}\label{eq: graph aug}
\mathbf{E'} = \mathcal{E}(\mathcal{G'}, \mathbf{E}^0), \mathbf{E''} = \mathcal{E}(\mathcal{G''}, \mathbf{E}^0),
\end{flalign}
\end{small}

\noindent where $\mathcal{E}(\cdot)$ denotes  graph encoder. Because nodes do not satisfy the IID assumption on the graph, random structure perturbation easy to destroys the intrinsic nature of the input graph, then can't fully make use of GCL for recommendation. Another is feature augmentation~\cite{simgcl}, which is illustrated in the lower part of Figure\ref{fig: intro}. Feature augmentation adds random noises into node embeddings, then generate contrastive representations with GNNs:
\begin{small}
\begin{flalign}\label{eq: emb aug}
\mathbf{E'} = \mathcal{E}(\mathbf{E}^0,\epsilon\delta'), \mathbf{E''} = \mathcal{E}(\mathbf{E}^0, \epsilon\delta''),
\end{flalign}
\end{small}

\begin{small}
\begin{figure*} [t]
  \begin{center}
    \includegraphics[width=1.0\textwidth]{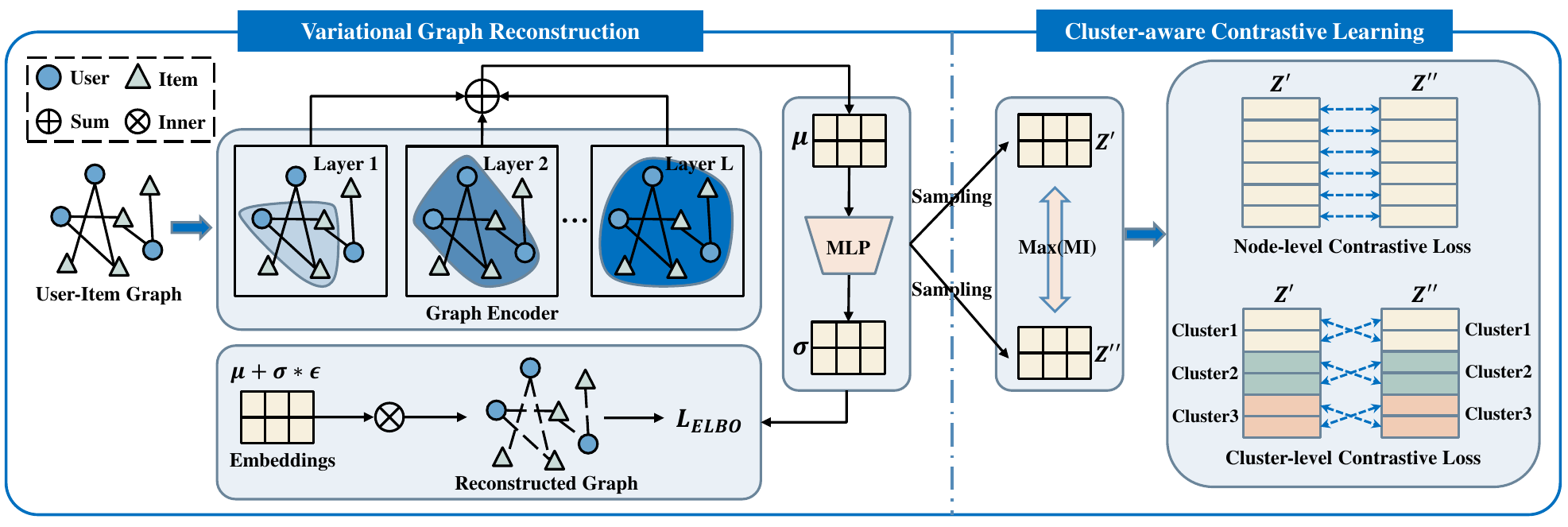}
  \end{center}
    \vspace{-0.35cm}
  \caption{An Illustration of our proposed \fullname~ framework, which consists of a variational graph reconstruction module and a cluster-aware contrastive learning module. The variational graph reconstruction module generates contrastive views by multiple samplings from the estimated distributions. The Cluster-aware contrastive learning module provides self-supervised signals, which include node-level and cluster-level contrastive objectives.}\label{fig:framework}
    \vspace{-0.35cm}
\end{figure*}
\end{small}

\noindent where $\delta', \delta''\sim U(0,1)$ are uniform noises, $\epsilon$ is the amplitude that controls noise scale. 
Although this noise-based augmentation is controllable and constrains the deviation, we argue that a fixed and generic $\epsilon$ is not generalized for nodes with unique characteristics. For example, user-item interactions usually perform the long-tail distribution, the head nodes have more supervision signals than tails, then a small $\epsilon$ maybe satisfy the tail nodes while not sufficient to the head nodes. The above flaws drive us to find a better graph augmentation that maintains graph information and is adaptive to each node.

\section{Methodology}
In this section, we present our proposed \fullname~ framework for recommendation. As shown in Figure \ref{fig:framework}, \shortname~ consists of two modules: a variational graph reconstruction module and a cluster-aware contrastive learning module. Specifically, we first use variational graph reconstruction to estimate the probability distribution of each node, then design cluster-aware twofold contrastive learning objectives to encourage the consistency of contrastive views which are generated by multiple samplings from the estimated distribution. Next, we introduce each component in detail.

\subsection{Variational Graph Reconstruction}
\textbf{VAE Brief.} Given the training data $\mathbf{X}=\{\mathbf{x}_i\}_{i=1}^n$, VAE assumes that each sample $\mathbf{x}_i$ is constructed from a generative process: $\mathbf{x}\sim p_\theta(\mathbf{x}|\mathbf{z})$. Thus, it's natural to maximize the likelihood function:
\begin{small}
    \begin{flalign} \label{eq: likelihood}
       log p(\mathbf{x}) = log\int p_\theta(\mathbf{x}|\mathbf{z})p(\mathbf{z})d\mathbf{z},
    \end{flalign}
\end{small}

\noindent where $p(\mathbf{z})$ is the prior distribution of latent variable $\mathbf{z}$. However, it's intractable to compute Eq.\eqref{eq: likelihood} because we don't know all possible latent variables $\mathbf{z}$. Thus, VAE adopts a variational inference technique and uses an inference model $q_\phi(\mathbf{z}|\mathbf{x})$ to approximate the posterior distribution $p_\theta\mathbf(\mathbf{x}|\mathbf{z})$. Then, VAE is optimized by minimizing the Evidence Lower Bound~(ELBO) based objective:
\begin{small}
    \begin{flalign}
        \mathcal{L}_{ELBO}=-\mathbb{E}_{\mathbf{z}\sim q_\phi(\mathbf{z}|\mathbf{x})}[log(p_\theta(\mathbf{x}|\mathbf{z}))] + KL[q_\phi(\mathbf{z}|\mathbf{x})||p(\mathbf{z})],
    \end{flalign}
\end{small}

\noindent where $q_\phi(\mathbf{z}|\mathbf{x})$ and $p_\theta(\mathbf{x}|\mathbf{z})$ also denote the encoder and decoder which are parameterized by neural networks. $KL[q_\phi(\mathbf{z}|\mathbf{x})||p(\mathbf{z})]$ is the Kullback-Leibler divergence between the approximate posterior $q_\phi(\mathbf{z}|\mathbf{x})$ and prior $p(\mathbf{z})$, which is used to constrain $q_\phi(\mathbf{z}|\mathbf{x})$ closer to the prior Gaussian distribution.

\textbf{Graph Inference.} Given the observed user-item interaction graph $\mathcal{G}=\{U\cup V, \mathbf{A}\}$, and initialized node embeddings $\mathbf{E}^0$. Graph inference aims to learn probability distributions $\mathbf{Z}$ which can reconstruct the input graph structure: $\mathbf{\hat{A}}\sim p_\theta(\mathbf{A}|\mathbf{Z})$. Same to VAE, we also adopt variational inference $q_\phi(\mathbf{Z}|\mathbf{A}, \mathbf{E}^0)=\prod_{i=0}^{M+N-1}q_\phi(\mathbf{z}_i|\mathbf{A},\mathbf{E}^0)$ to approximate the posterior $p_\theta(\mathbf{A}|\mathbf{Z})$. To be specific, we encode each node $i$ into a multi-variate Gaussian distribution $q_\phi(\mathbf{z}_i|\mathbf{A}, \mathbf{E}^0) =\mathcal{N}(\mathbf{z}_i|\mathbf{\mu}_\phi(i), diag(\mathbf{\sigma}_\phi^2(i)))$, where $\mathbf{\mu}_\phi(i)$ and $\mathbf{\sigma}_\phi^2(i)$ denote the mean and variance of node $i$'s distribution, respectively. To better exploit high-order user-item graph structure, we adopt GNNs to estimate the parameters of node distributions:
\begin{small}
    \begin{flalign}
        \mathbf{\mu} = GNN(\mathbf{A},\mathbf{E}^0,\phi_\mu), \mathbf{\sigma} = GNN(\mathbf{A},\mathbf{E}^0,\phi_\sigma),
    \end{flalign}
\end{small}

\noindent where $\phi_\mu$ and $\phi_\sigma$ denote learnable parameters on graph inference. Following the previous research on graph-based collaborative filtering, we select LightGCN~\cite{LightGCN} as the encoder to deploy the above graph inference process. For each node $i$, the corresponding means are updated as follows:
\begin{small}
    \begin{flalign} \label{eq: graph inference}
        \mathbf{\mu}_i^l &=\sum_{j\in \mathcal{N}_i}\frac{1}{\sqrt{|\mathcal{N}_i|} \sqrt{|\mathcal{N}_j|}} \mathbf{\mu}_i^{l-1},
    \end{flalign}
\end{small}

\noindent where $\mathbf{\mu}_i^l$ and $\mathbf{\mu}_i^{l-1}$ are corresponding means on $l^{th}$ and ${(l-1)}^{th}$ graph convolution layer, $\mathcal{N}_i$ and $\mathcal{N}_j$ denote the connected neighbors for node $i$ and node $j$. We initialize the means $\mathbf{\mu}^0=\mathbf{E}^0$.
When stacking $L$ graph convolution layers, we have $L+1$ outputs $[\mu^0, \mu^1, ..., \mu^{L}]$, then we fuse all layers' outputs and compute the means and variances as follows:
\begin{small}
    \begin{flalign} \label{eq: para_mean_var}
        \mathbf{\mu} = \frac{1}{L}\sum_{l=1}^L\mu^l,
        \mathbf{\sigma} = MLP(\mathbf{\mu}),
    \end{flalign}
\end{small}

\noindent where the variances are learned from an MLP, which feeds the means as the input. In practice, we find that one-layer MLP achieves the best performance, then $\mathbf{\sigma}=exp(\mu \mathbf{W} + \mathbf{b})$, where $\mathbf{W} \in \mathbb{R}^{d\times d}$ and $\mathbf{b} \in \mathbb{R}^d$ are two learnable parameters.
After obtaining the mean and variance of the approximate posterior, we generate the latent representation $\mathbf{z}_i$ by sampling from $\mathcal{N}(\mu_i, \sigma^2_i)$. However, it can not be directed optimized because the sampling process is non-differentiable. We employ the reparameterization trick instead of the sampling process~\cite{kingma2013vae}: 
\begin{small}
    \begin{flalign} \label{eq: repara}
        \mathbf{z}_i=\mathbf{\mu}_i+\mathbf{\sigma}_i\cdot\varepsilon,
    \end{flalign}
\end{small}

\noindent where $\varepsilon \sim \mathcal{N}(0, \mathbf{I})$ is a normal Gaussian noise. 

\textbf{Graph Generation.}
After estimating the probability distribution of the latent variables $\mathbf{Z}$, the objective of graph generation is to reconstruct the original user-item graph:
\begin{small}
    \begin{flalign}
        p(\mathbf{A}|\mathbf{Z}) = \prod_{i=0}^{M+N-1}\prod_{j=0}^{M+N-1}p(\mathbf{A}_{ij}|\mathbf{z}_i,
        \mathbf{z}_j).
    \end{flalign}
\end{small}

\noindent There are many choices to realize the graph generation process, such as inner product, factorization machine, and neural networks. As suggested in~\cite{kipf2016vgae}, we use an inner product to compute the propensity score that node $i$ connected with node $j$:
\begin{small}
    \begin{flalign} \label{eq: graph generation}
       p(\mathbf{A}_{ij}=1|\mathbf{z}_i,\mathbf{z}_j) =\sigma(\mathbf{z}_i^T\mathbf{z}_j), 
    \end{flalign}
\end{small}

\noindent where $\sigma(\cdot)$ is the sigmoid function.

\subsection{Cluster-aware Contrastive Learning}
\textbf{Contrastive View Construction.} Given the estimated probability distribution of latent representation $\mathbf{Z}\sim \mathcal{N}(\mathbf{\mu},\mathbf{\sigma}^2)$, we introduce a novel contrastive learning paradigm based on the estimated distribution. Different from previous GCL-based recommendation methods~\cite{sgl,simgcl}, we construct contrastive views through multiple samplings from the estimated distribution instead of data augmentation. Specifically, for each node $i$, we generate contrastive representations $\mathbf{z}'$ and $\mathbf{z}''$ as follows:
\begin{small}
    \begin{flalign}\label{eq: samplings}
        \mathbf{z}_i'&= \mathbf{\mu}_i + \mathbf{\sigma}_i\cdot\varepsilon', \\
        \mathbf{z}_i''&=\mathbf{\mu}_i + \mathbf{\sigma}_i\cdot\varepsilon'',
    \end{flalign}
\end{small}

\noindent where $\varepsilon',\varepsilon'' \sim \mathcal{N}(0, \mathbf{I})$ are two random normal noise. Compared to structure or feature augmentations, our method is more efficient and effective for contrastive view construction. Firstly, all contrastive representations are sampled from the estimated distributions, which can well reconstruct the input graph without any information distortion. Secondly, the estimated variances are tailored to each node, which can be adaptive to regulate the scale of contrastive loss.

\textbf{Node-level Contrastive Loss.} 
After constructing contrastive views of each node, we maximize the mutual information to provide self-supervised signals to improve recommendation performance. Considering that similar nodes are closer in the representation, we propose cluster-aware twofold contrastive objectives for optimization: a node-level contrastive loss and a cluster-level contrastive loss. Among them, node-level contrastive loss encourages consistency of contrastive views for each node, and cluster-level contrastive loss encourages consistency of contrastive views of nodes in a cluster. The objective of node-level contrastive learning is $\mathcal{L}_N=\mathcal{L}_N^U+\mathcal{L}_N^V$, where $\mathcal{L}_N^U$ and $\mathcal{L}_N^V$ denote user side and item side losses:
\begin{small}
\begin{flalign} \label{eq: node-level}
    \mathcal{L}_{N}^U &= \sum\limits_{a \in \mathcal{B}_u}
    -log \frac{exp({\mathbf{z}'_a}^T\mathbf{z}''_a/\tau_1)}{\sum\limits_{b \in \mathcal{B}_u}
    exp({\mathbf{z}'_a}^T\mathbf{z}''_b/\tau_1)}, \\
    \mathcal{L}_N^I &= \sum\limits_{i \in \mathcal{B}_i} 
    -log \frac{exp({\mathbf{z}'_i}^T\mathbf{z}''_i/\tau_1)}
    {\sum\limits_{j \in \mathcal{B}_i}
    exp({\mathbf{z}'_i}^T\mathbf{z}''_j/\tau_1)},
\end{flalign}
\end{small}

\noindent where $\tau_1$ is the contrastive temperature, $\mathcal{B}_u$ and $\mathcal{B}_i$ denote users and items in a batch training data.

\textbf{Cluster-level Contrastive Loss.}
Considering the similarity of the estimated distributions of nodes, we design cluster-level contrastive loss to further distinguish the positive and negative contrastive pairs in batch training data. Overall, our aim is to maximize the consistency of node pairs with the same cluster and minimize the consistency of node pairs with different clusters. 
Suppose there are $K_u$ user prototypes $\mathbf{C}^u \in \mathbb{R}^{d \times K_u}$ and $K_i$ item cluster prototypes $\mathbf{C}^i \in \mathbb{R}^{d \times K_i}$, we use $p(c_k^u|z_a)$ to denote the conditional probability that user $a$ belongs to $k^{th}$ user cluster, and $p(c_h^i|z_i)$ denote the conditional probability that item $i$ belongs to $h^{th}$ item cluster. Given the estimated distributions as input, we implement the clustering process by the K-Means algorithm~\cite{hartigan1979algorithm}. Then, we compute the probability that two users~(items) are assigned to the same prototype:
\begin{small}
\begin{flalign}
    p(a,b) &= \sum_{k=0}^{K_u-1}p(\mathbf{c}_k^u|\mathbf{z}_a)p(\mathbf{c}_k^u|\mathbf{z}_b), \\
    p(i,j) &= \sum_{h=0}^{K_i-1}p(\mathbf{c}_h^i|\mathbf{z}_i)p(\mathbf{c}_h^i|\mathbf{z}_j),
\end{flalign}
\end{small}

\noindent where $p(a,b)$ denote the probability that user $a$ and user $b$ belong to the same cluster, and $p(i,j)$ denote the probability that item $i$ and item $j$ belong to the same cluster. Next, we present the cluster-level contrastive loss $\mathcal{L}_C=\mathcal{L}_C^U+\mathcal{L}_C^I$, where $\mathcal{L}_C^U$ and $\mathcal{L}_C^I$ denote user side and item side losses:
\begin{small}
\begin{flalign}\label{eq: cluster_level loss}
    \mathcal{L}_{C}^{U} &=\sum\limits_{a \in \mathcal{B}_u}
    \frac{-1}{SP(a)}log(\frac{\sum\limits_{b \in {\mathcal{B}_u, b!=a}}p(a,b)exp(\mathbf{z'}_a^T\mathbf{z''}_b/{\tau}_2)}{\sum\limits_{b \in \mathcal{B}_u,b!=a}exp(\mathbf{z'}_a^T\mathbf{z''}_b/{\tau}_2)}), \\
    \mathcal{L}_{C}^{I} &= \sum\limits_{i \in \mathcal{B}_i}\frac{-1}{SP(i)}
    log(\frac{\sum\limits_{j \in \mathcal{B}_i,j!=i}p(i,j)exp(\mathbf{z'}_i^T\mathbf{z''}_j/{\tau}_2)}{\sum\limits_{j \in \mathcal{B}_i,j!=i}exp(\mathbf{z'}_i^T\mathbf{z''}_j/{\tau}_2)}),
\end{flalign} 
\end{small}

\noindent where $SP(a)=\sum\limits_{b \in \mathcal{B}_u, b!=a} p(a,b)$ and $SP(i)=\sum\limits_{j \in \mathcal{B}_i, j!=i} p(i,j)$, $\tau_2$ is the temperature to control the mining scale of hard negatives. The final contrastive loss is the weighted sum of the node-level loss and the cluster-level contrastive loss:
\begin{small}
    \begin{flalign}
        \mathcal{L}_{cl}=\mathcal{L}_N+\gamma\mathcal{L}_C,
    \end{flalign}
\end{small}

\noindent where $\gamma$ is the coefficient to balance two level contrastive losses.

\begin{algorithm} [t]
\renewcommand{\algorithmicrequire}{\textbf{Input:}}
\renewcommand\algorithmicensure {\textbf{Output:}}
\caption{\small{The Algorithm of \shortname}}\label{alg: vgcl}
\begin{algorithmic}[1]
\REQUIRE user-item bipartite graph $\mathcal{G}$;  ~~\\
\ENSURE Parameters $\Theta_{GNN}=\mathbf{E}^0$ and $\Theta_{MLP}=[\mathbf{W}, \mathbf{b}]$; ~~\\

\STATE Randomly initialize parameters $\Theta_{GNN}$ and $\Theta_{MLP}$ ; \\
\WHILE{not converged}
    \STATE Sample a batch of training data;
    \STATE Calculate graph inference parameters $\mu$ and $\sigma$~(Eq.\eqref{eq: graph inference} to Eq.\eqref{eq: para_mean_var}); 
    \STATE Estimate node distribution $\mathbf{Z}$ by parameterization (Eq.\eqref{eq: repara});
    \STATE Generate contrastive instances $\mathbf{Z}'$ and $\mathbf{Z}''$ by multiple samplings ~(Eq.\eqref{eq: samplings}, Eq.(18));
    \STATE Compute prototypes $\mathbf{C}_u$ and $\mathbf{C}_v$ based on K-Means clustering algorithm;
    \STATE Compute node-level contrastive loss $\mathcal{L}_N$~(Eq.\eqref{eq: node-level});
    \STATE Compute cluster-level contrastive loss $\mathcal{L}_C$~(Eq.(23), Eq.(24));
    \STATE Compute variational graph reconstruction loss $\mathcal{L}_{ELBO}$~(Eq.\eqref{eq: loss_elbo});
    \STATE Update all parameters according to ~(Eq.\eqref{eq: objectives});
\ENDWHILE
\STATE Return $\Theta_{GNN}=\mathbf{E}^0$ and $\Theta_{MLP}=[\mathbf{W}, \mathbf{b}]$.
\end{algorithmic}
\end{algorithm}

\subsection{Model Optimization.} 
For the variational graph reconstruction part, we optimize the parameters of graph inference and graph generation with ELBO:
\begin{small}
    \begin{flalign} \label{eq: loss_elbo}
        \mathcal{L}_{ELBO}=-\mathbb{E}_{\mathbf{Z}\sim q_\phi(\mathbf{Z}|\mathbf{A},\mathbf{E}^0)}[log(p_\theta(\mathbf{A}|\mathbf{Z}))]
        + KL[q_\phi(\mathbf{Z}|\mathbf{A},\mathbf{E}^0)||p(\mathbf{Z})].
    \end{flalign} 
\end{small}

\noindent Among them, the first term is the reconstruction error between the original graph and the generated graph. We employ a pairwise learning strategy to minimize the reconstruction error:
\begin{small}
    \begin{flalign}
\mathbb{E}_{\mathbf{Z}\sim q_\phi(\mathbf{Z}|\mathbf{A},\mathbf{E}^0)}[log(p_\theta(\mathbf{A}|\mathbf{Z}))]=
\sum_{a=0}^{M-1}\sum\limits_{(i,j)\in D_a}-log\sigma(\hat{r}_{ai}-\hat{r}_{aj})),
    \end{flalign} 
\end{small}

\noindent {\small$D_a=\{(i,j)|i\in R_a\!\wedge\!j\not\in R_a\}$} denotes the pairwise training data for user $a$. {\small$R_a$} represents the item set that user $a$ has interacted. Overall,
we optimize the proposed \shortname~ with a multi-task learning framework:
\begin{small}
    \begin{flalign}\label{eq: objectives}
    min \mathcal{L}=\mathcal{L}_{ELBO} + \alpha \mathcal{L}_{cl} + \lambda||\mathbf{E}^0||^2, 
    \end{flalign} 
\end{small}

\noindent where $\alpha$ is the balance parameter of contrastive loss, and $\lambda$ is the 
regularization coefficient. After the model training process, we use Eq.\eqref{eq: graph generation} to predict the unknown preferences for the recommendation.

\subsection{Model Analysis}
\textbf{Space Complexity.} As shown in Algorithm 1, the model parameters are composed of two parts: node embeddings $\mathbf{E}^0$ and MLP parameters $\mathbf{W}, \mathbf{b}$. Compared to traditional embedding-based collaborative filtering, the additional parameters only have $\mathbf{W}, \mathbf{b}$ which are shared among all nodes. So the additional storage space is very small and can be neglected. 

\noindent \textbf{Time Complexity.} We compare the time complexity of \shortname~ with other GCL-based recommendation methods based on data augmentation. Let $|E|$ denote the edge number of the graph, $d$ be the embedding size, and $S$ denote the average neighbor number. For the graph convolution part, \shortname~ costs $\mathcal{O}(2|E|dS)$, where $2|E|$ denotes the number of non-zero elements on the adjacent matrix. However, SGL and SimGCL need to repeat graph convolution three times, which generates main embeddings for recommendation and two auxiliary embeddings for contrastive learning. Therefore, SGL and SimGCL all cost $\mathcal{O}(6|E|dS)$ while \shortname~ only need $\mathcal{O}(2|E|dS)$.
For the contrastive learning part, \shortname~ additionally has a clustering process, we implement the K-means clustering algorithm with Faiss-GPU~\footnote{https://faiss.ai/}, and the time cost can be neglected compared to model learning in practice. Therefore, \shortname~ is more time-efficient than current GCL-based recommendation methods based on data augmentation.

\section{Experiments}
\subsection{Experimental Settings}
\subsubsection{Datasets}
To compare the recommendation performance of our \shortname~ with other state-of-the-art models, we select three benchmarks to conduct the empirical analysis: Douban-Book~\cite{simgcl}, Dianping~\cite{wu2020diffnet++} and Movielens-25M~\cite{harper2015movielens}. For Movielens-25M, we convert ratings equal to 5 as positive feedback, and other ratings as negative feedback. We filter users with less than 10 interactions for all datasets, and randomly sample 80\% interactions as training data, and the remaining 20\% as test data. The statistics of three datasets are summarized in Table~\ref{tab: statistics}.

\begin{table}[t]
    \centering
	\setlength{\belowcaptionskip}{5pt} %
	\caption{The statistics of three datasets.}\label{tab: statistics}
	 \vspace{-0.2cm}
	{{
    \begin{tabular}{c|c|c|c|c}
    \hline
    Datasets & Users & Items & Interactions & Density\\ \hline
    Douban-Book & 13,024 & 22,347 & 792,062 & 0.272\% \\  \hline
    Dianping & 59,426 & 10,224 & 934,334 & 0.154\%\\ \hline
    Movielens-25M & 92,901 & 8,826 & 2,605,952 & 0.318\% \\ \hline
    \end{tabular}}}
\end{table}

\begin{table*}[t]\centering
\centering
\caption{Recommendation performances on three datasets. The best-performing model on each dataset and metrics are highlighted in bold, and the second-best model is underlined.}
\label{tab: overall performance}
\scalebox{1.06}{
\begin{tabular}{|l|c|c|c|c|c|c|c|c|c|c|c|c|}
\hline 
 & \multicolumn{4}{c|}{Douban-Book}       
 & \multicolumn{4}{c|}{Dianping}
 & \multicolumn{4}{c|}{Movielens-25M}\\ \cline{2-13}
\multirow{-2}{*}{Models} & R@10 & N@10 & R@20 & N@20 & R@10 & N@10 & R@20 & N@20 & R@10 & N@10 & R@20 & N@20\\ \hline
BPR-MF  & 0.0869 & 0.0949 & 0.1296 & 0.1045 & 0.0572 & 0.0443 & 0.0934 & 0.0557 & 0.2152 & 0.2011 & 0.3163 & 0.2343 \\ \hline
LightGCN  & 0.1042 & 0.1195 & 0.1516 & 0.1278 & 0.0679 & 0.0536 & 0.1076 & 0.0660 & 0.2258 & 0.2192 & 0.3263 & 0.2509 \\ \hline
Multi-VAE & 0.0941 & 0.1073 & 0.1376 & 0.1155 & 0.0645 & 0.0508 & 0.1046 & 0.0632 & 0.2188 & 0.2101 & 0.3185 & 0.2418 \\ \hline
CVGA & 0.1058 & 0.1305 & 0.1492 & 0.1359 & 0.0719 & 0.0562 & 0.1128 & 0.0690 & 0.2390 & 0.2306 & 0.3454 & 0.2641 \\ \hline
SGL-ED & 0.1103 & 0.1357 & 0.1551 & 0.1419 & 0.0719 & 0.0560 & 0.1111 & 0.0686 & 0.2298 & 0.2239 & 0.3274 & 0.2541  \\ \hline
NCL  & 0.1121 & 0.1377 & 0.1576 & 0.1439 & 0.0727 & 0.0571 & 0.1124 & 0.0701 & 0.2281 & 0.2222 & 0.3274 & 0.2531\\ \hline
SimGCL  & \underline{0.1218} & \underline{0.1470} & \underline{0.1731} & \underline{0.1540} & \underline{0.0768} & \underline{0.0606} & \underline{0.1208} & \underline{0.0743} & \underline{0.2428} & \underline{0.2356} & \underline{0.3491} & \underline{0.2690}\\ \hline
\textbf{VGCL}  & \textbf{0.1283} & \textbf{0.1564} & \textbf{0.1829} & \textbf{0.1638} & \textbf{0.0778} & \textbf{0.0616} & \textbf{0.1234} & \textbf{0.0757} & \textbf{0.2463} & \textbf{0.2400} & \textbf{0.3507} & \textbf{0.2725}\\ \hline
\end{tabular}}
\end{table*}

\subsubsection{Baselines and Evaluation Metrics}
We compare our model with the following baselines, including matrix factorization based method: BPR-MF~\cite{UAI2009BPR},
graph based method: LightGCN~\cite{LightGCN}, VAE based methods: Multi-VAE~\cite{Multi-VAE}, CVGA~\cite{zhang2022revisiting}, and graph contrastive learning based methods: SGL~\cite{sgl}, NCL~\cite{ncl}, SimGCL~\cite{simgcl}.

We employ two widely used metrics: Recall@N and NDCG@N to evaluate all recommendation models. Specifically, Recall@N measures the percentage of recalled items on the Top-N ranking list, while NDCG@N further assigns higher scores to the top-ranked items. To avoid selection bias in the test stage, we use the full-ranking strategy~\cite{zhao2020revisiting} that views all non-interacted items as candidates. All metrics are reported with average values with 5 times repeated experiments.

\subsubsection{Parameter Settings}
We implement our \shortname~ model and all baselines with Tensorflow\footnote{https://www.tensorflow.org}. We initialize all models parameter with a Gaussian distribution with a mean value of 0 and a standard variance of 0.01, embedding size is fixed to 64. We use Adam as the optimizer for model optimization, and the learning rate is 0.001. The batch size is 2048 for the Douban-Book and Dianping datasets and 4096 for the Movielens-25M dataset. For our \shortname~ model, we turn the contrastive temperature $\tau$ in $[0.10, 0.25]$, contrastive regularization coefficient $\lambda$ in $[0.01, 0.05, 0.1, 0.2, 0.5, 1.0]$, and clustering number $k_1, k_2$ in $[100, 1000]$. Besides, we carefully search the best parameter of $\gamma$, and find \shortname~ achieves the best performance when $\gamma=0.4$ on Douban-Book, $\gamma=0.5$ on Dianping dataset, and $\gamma=1.0$ on Movielens-25M dataset. As we employ the pairwise learning strategy for graph reconstruction, we randomly select one unobserved item as a candidate negative sample to compose triple data for model training. For all baselines, we search the parameters carefully for fair comparisons. We repeat all experiments 5 times and report the average results.

\begin{table*}[t]\centering
	\begin{small}
		\centering
		\setlength{\belowcaptionskip}{5pt}
		\caption{Ablation study of VGCL, VGCL-w/o C denotes without cluster-level contrastive loss and VGCL-w/o V denotes without the variational graph reconstruction part.}\label{tab:ablation study}
		\vspace{-5pt}
        \scalebox{1.16}{
		\begin{tabular}{|c|c|c|c|c|c|c|} \hline
		\multirow{2}{*}{Models} &
		\multicolumn{2}{|c|}{Douban-Book}& \multicolumn{2}{|c|}{Dianping} & \multicolumn{2}{|c|}{Movielens-25M} \\
		\cline{2-7}
	&R@20&N@20&R@20&N@20&R@20&N@20\\ \hline
        LightGCN & 0.1512(-) & 0.1271(-) & 0.1076(-) & 0.0660(-) & 0.3263(-) & 0.2509(-) \\ \hline
        SimGCL & 0.1731(14.48\%) & 0.1540(+21.16\%) & 0.1208(+12.27\%) & 0.0743(+12.58\%) & 0.3491(+6.99\%) & 0.2690(+7.21\%) \\ \hline
        VGCL-w/o C & 0.1776(+17.46\%) & 0.1575(+23.92\%) & 0.1222(+13.57\%) & 0.0750(+13.64\%) & 0.3477(+6.56\%) & 0.2705(+7.81\%) \\ \hline
        VGCL-w/o V & 0.1722(+13.89\%) & 0.1547(+21.72\%) & 0.1218(+13.20\%) & 0.0746(+13.03\%) & 0.3493(+7.05\%) & 0.2702(+7.69\%) \\ \hline
        \textbf{VGCL} & \textbf{0.1829(+20.97\%)} & \textbf{0.1638(+28.87\%)} & \textbf{0.1233(+14.59\%)} & \textbf{0.0756(+14.55\%)} & \textbf{0.3507(+7.48\%)} & \textbf{0.2725(+8.61\%)} \\ \hline 
		\end{tabular}}
	\end{small}
\end{table*}

\subsection{Overall Performance Comparisons}
As shown in Table~\ref{tab: overall performance}, we compare our model with other baselines on three datasets. We have the following observations:
\begin{itemize}
    \item Our proposed \shortname~ consistently outperforms all baselines under different settings. Specifically, \shortname~ improves LightGCN $w.r.t$ NDCG@20 by 28.17\%, 14.70\% and 8.61\% on Douban-Book, Dianping and Movielens-25M dataset, respectively. Compared to the strongest baseline~(SimGCL), \shortname~ also achieves better performance, e.g., about 6.36\% performance improvement of NDCG@20 on the Douban-Book dataset. Besides, we find that \shortname~ achieves higher improvements on the small-length ranking task, which is more suitable for real-world recommendation scenarios. Extensive empirical studies verify the effectiveness of the proposed \shortname~, which benefits from combining the strength of generative and contrastive graph learning for recommendation.
    \item Graph-based methods achieve better performance than their counterparts, which shows the superiority that capturing users' preferences by modeling high-order user-item graph structure. To be specific, LightGCN always outperforms BPR and CVGA consistently outperforms Multi-VAE, which proves that graph learning can effectively capture the high-order user-item interaction signals to improve recommendation performance, whether in embedding-based or VAE-based recommendation methods.
    \item All GCL-based methods~(SGL, NCL, SimGCL) significantly improve LightGCN on three datasets. It verifies the effectiveness of incorporating self-supervised learning into collaborative filtering. SimGCL achieves the best performance among these baselines, demonstrating that feature augmentation is more suitable for collaborative filtering than structure augmentation, which can maintain sufficient invariants of the original graph. It's worth noting that, our method also can be regarded as feature augmentation, but we rely on multiple samplings from the estimated distribution and the scales of augmentations are adaptive to different nodes. Therefore, \shortname~ achieves better performance compared to SimGCL.
\end{itemize}

\vspace{-0.2cm}
\subsection{Ablation Study}
To exploit the effectiveness of each component of the proposed \shortname~, we conduct the ablation study on three datasets. As shown in Table~\ref{tab:ablation study}, we compare \shortname~ and corresponding variants on Top-20 recommendation performances. VGCL-w/o C denotes that remove the cluster-level contrastive loss of \shortname~, we only use the general node-level contrastive loss. 
VGCL-w/o V denotes that remove the variational graph reconstruction part of \shortname~, then we use feature augmentation the same as SimGCL to generate contrastive views. From Table~\ref{tab:ablation study}, we observe that VGCL-C consistently improves SimGCL on three datasets, which verifies that the proposed variational graph reconstruction module can provide better contrastive views for contrastive learning. Besides, VGCL-V also shows better performances than SimGCL, it demonstrates the effectiveness of cluster-aware contrastive pair sampling on contrastive learning. Finally, \shortname~ consistently outperforms two variants, demonstrating the effectiveness of combining the variational graph reconstruction and cluster-aware sampling strategy. Based on the above analysis, we can draw the conclusion that variational graph reconstruction can provide better contrastive views than simple data augmentation and cluster-aware sampling is better than random sampling for contrastive learning. All of the proposed modules are beneficial to GCL-based recommendations.

\begin{figure} [t]
  \begin{center}
      \subfigure[Douban-Book Dataset]{
      \includegraphics[width=40mm]{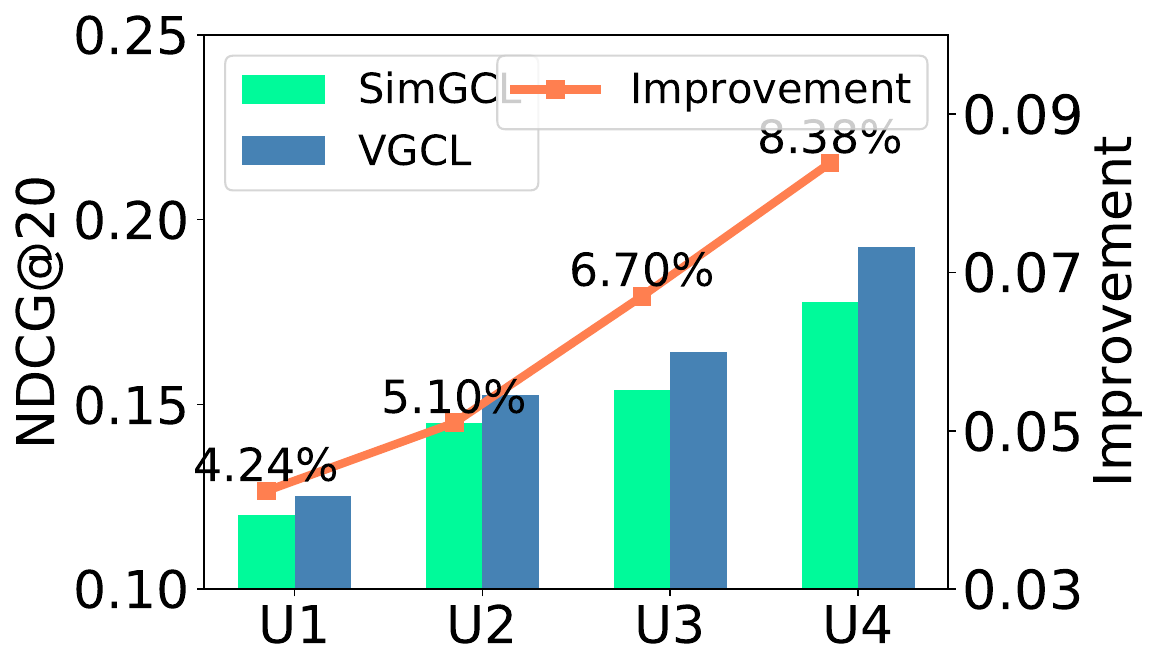}
      \includegraphics[width=40mm]{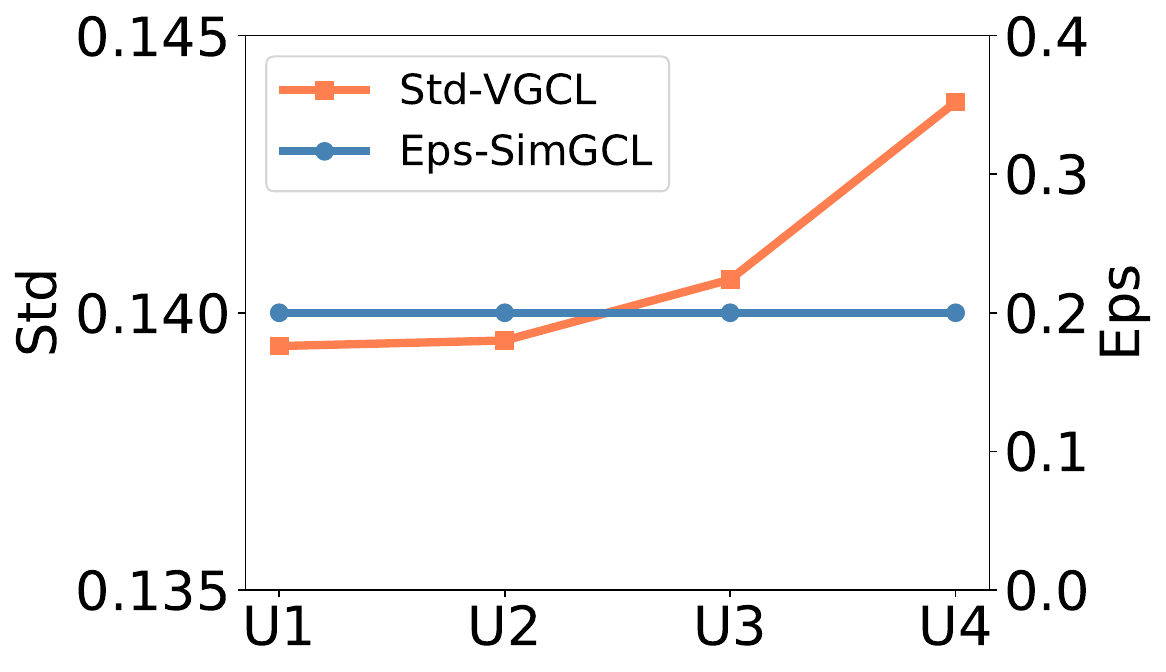}}
      \vspace{-0.3cm}
      \subfigure[Dianping Dataset]{
      \includegraphics[width=40mm]{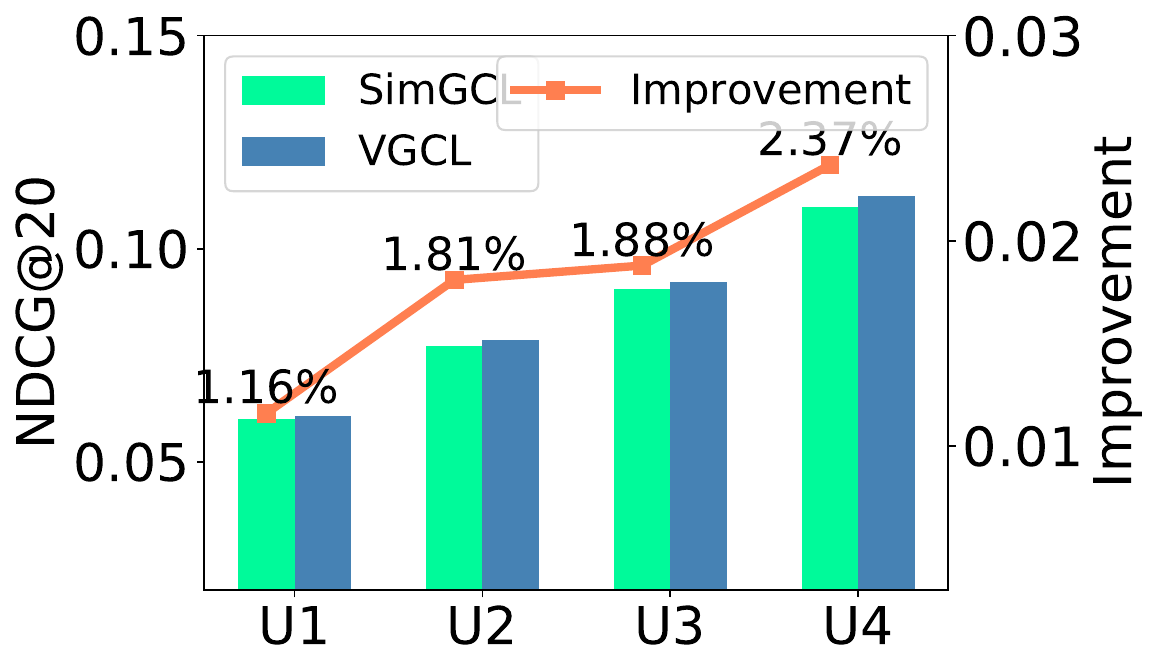}
      \includegraphics[width=40mm]{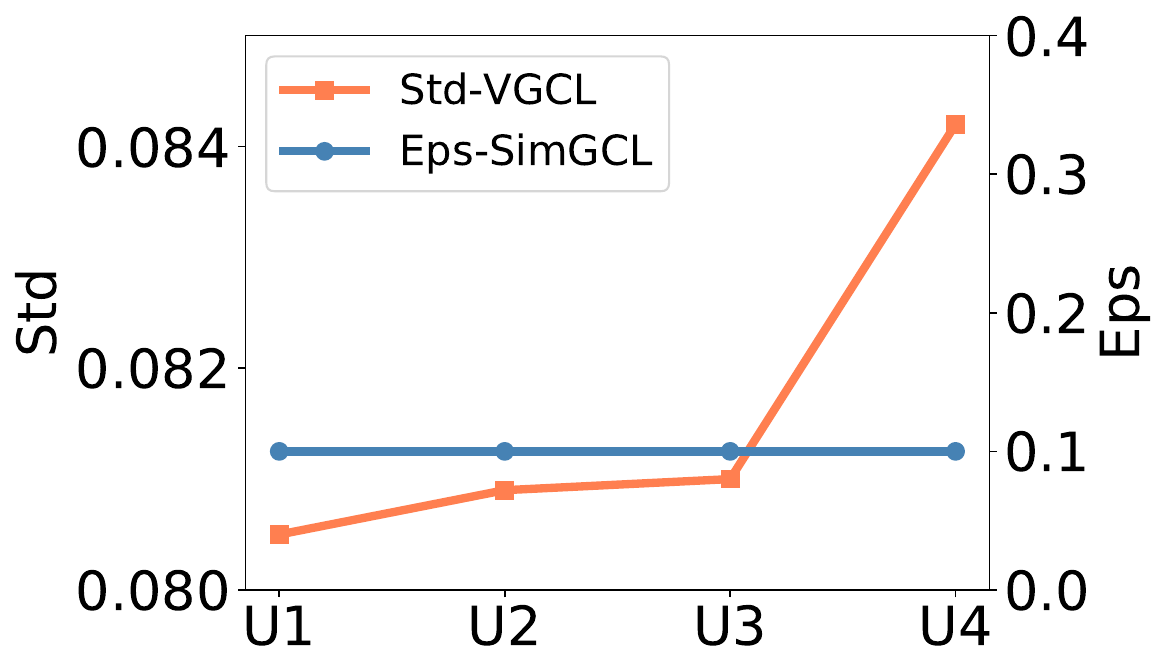}}
  \end{center}
  \vspace{-0.3cm}
  \caption{\small{Performance comparisons under different user groups.}} 
  \label{fig: group improvement}
  \vspace{-0.3cm}
\end{figure}

\begin{figure*} [th]
  \begin{center}
      \subfigure[$\tau$ on Douban-Book]{
      \includegraphics[width=42mm]{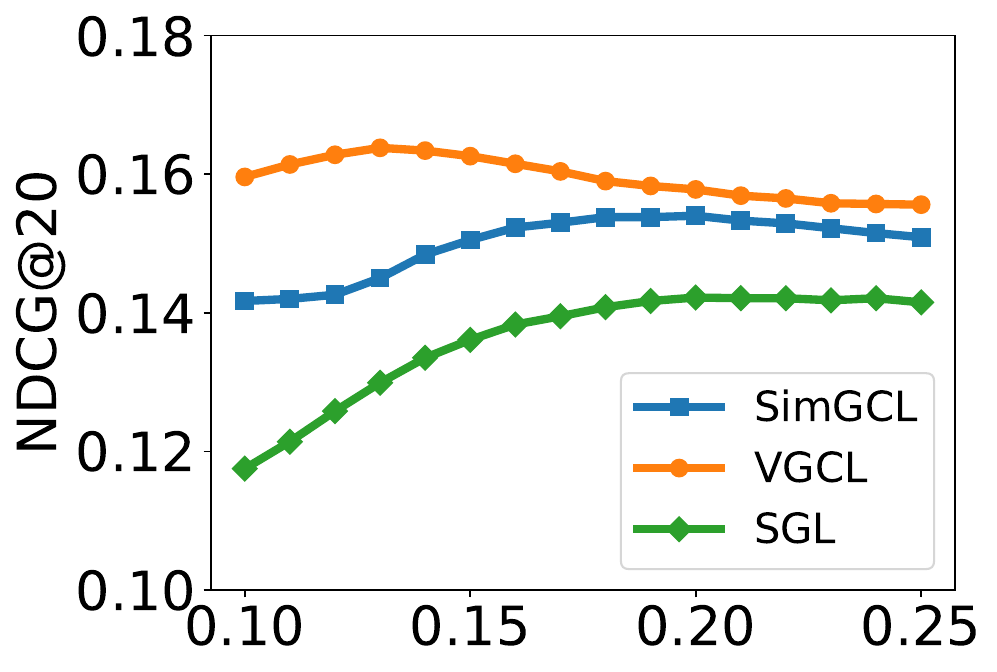}}
      \subfigure[$\tau$ on Dianping]{
      \includegraphics[width=42mm]{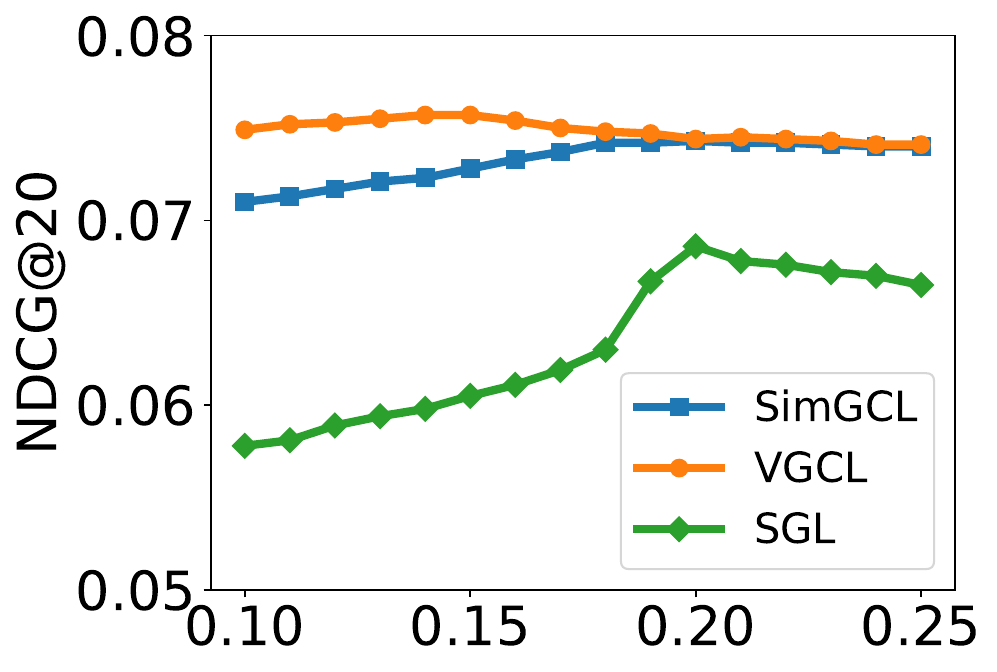}}
      \subfigure[$K_u, K_i$ on Douban-Book]{
      \includegraphics[width=44mm]{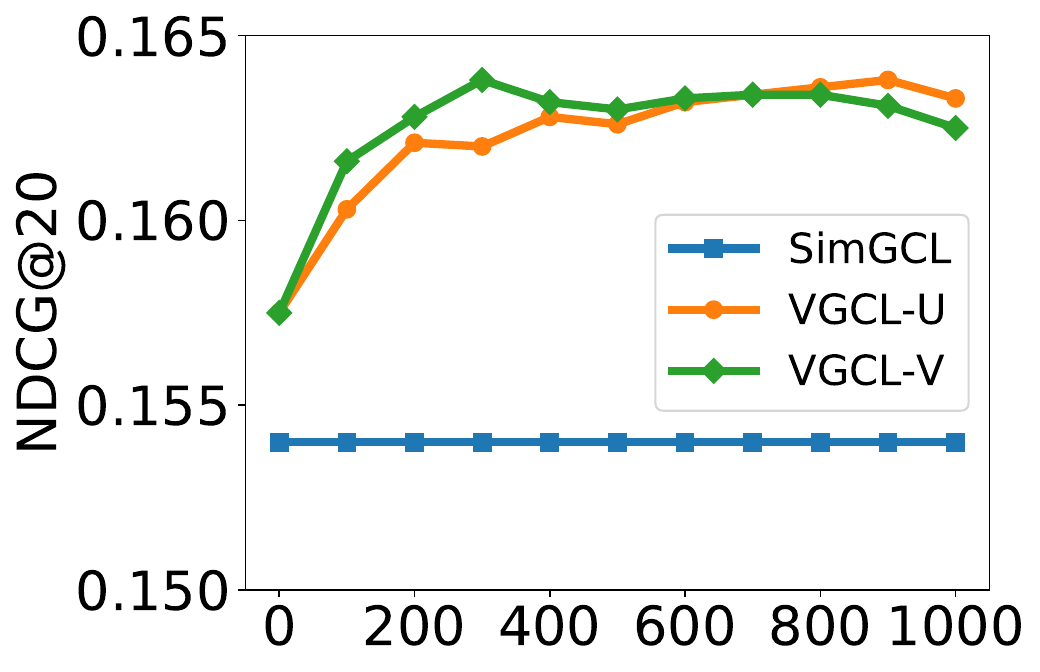}}
      \subfigure[$K_u, K_i$ on Dianping]{
      \includegraphics[width=44mm]{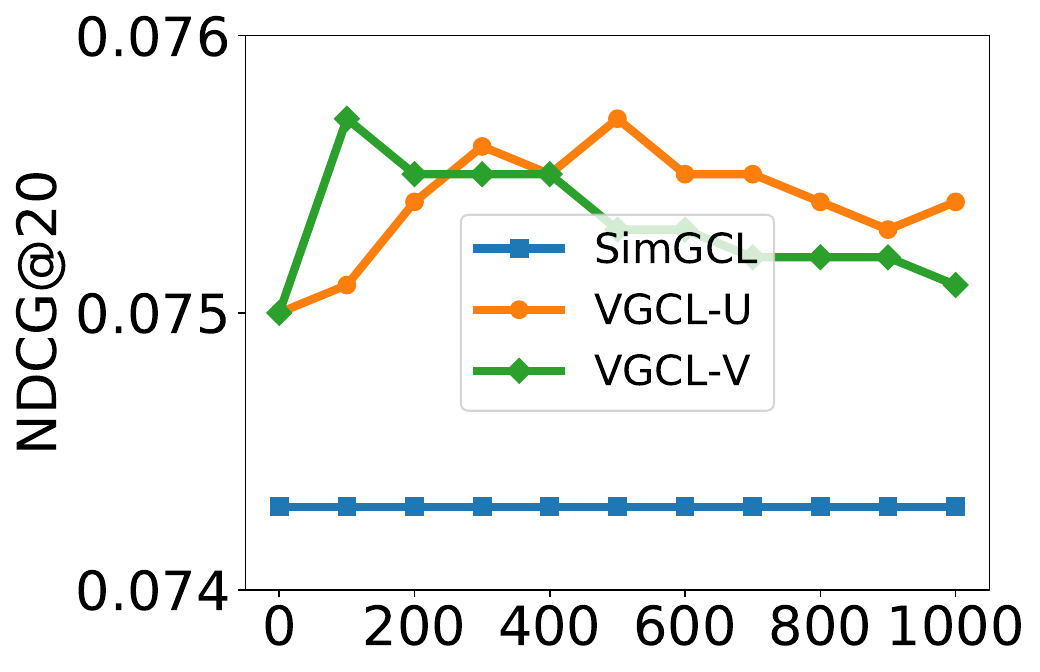}}
  \end{center}
  \vspace{-0.3cm}
  \caption{\small{Performance comparisons w.r.t different temperature $\tau$, prototype number $K_u$ and $K_i$.}} 
  \label{fig: temp_and_prototype}
  \vspace{-0.2cm}
\end{figure*}

\subsection{Investigation of the Estimated Distribution}
As we introduced in methodology, our proposed \shortname~ can adaptively learn variances for various nodes. To investigate the effect of personalized variances, we conduct comparisons of different user groups. Specifically, we first split all users into 4 groups according to their interactions, then analyze recommendation performances under different user groups. Figure\ref{fig: group improvement} illustrates NDCG@20 values of various groups on Douban-Book and Dianping datasets. We observe that all models show better performances in the denser user group, which conforms to the intuition of CF. Besides, our proposed \shortname~ achieves better performances on all user groups, demonstrating that \shortname~ is general to users with different interactions. Further, we plot the relative improvements that \shortname~ over SimGCL on Figure\ref{fig: group improvement}. We find that \shortname~ achieves a more significant improvement in the denser groups, e.g., 8.4\% improvement in U4 while 4.2\% improvement in U1 on the Douban-Book dataset. To exploit this phenomenon, we compared the standard variances of the estimated distribution of different users. From the right part of Figure\ref{fig: group improvement}, we can observe that the inferred standard variances vary from each group, and increase by group ID. Compared with SimGCL which set fixed eps~(noise scale) for all users, our method can learn personalized contrastive scales for different users. What's more, \shortname~ can adaptively learn larger variances to those users with amounts of interactions, it's important to provide sufficient self-supervised signals to improve recommendation performance. Experimental results effectively demonstrate the effectiveness of our proposed adaptive contrastive objectives.

\subsection{Hyper-Parameter Sensitivities}
In this part, we analyze the impact of hyper-parameters in \shortname~. We first exploit the effect of temperature $\tau$, which plays an important role in contrastive learning. Next, we investigate the influence of graph inference layer $L$. Finally, we study the impact of clustering prototype numbers $K_u, K_v$ and contrastive loss weights $\alpha$ and $\gamma$. 

\textbf{Effect of Graph Inference Layer $L$.}
To exploit the effect of different graph inference layers, we search the parameter $L$ in the range of $\{1,2,3,4\}$. As shown in Table~\ref{tab:inference layer}, we compare experimental results of different graph inference layers on Douban-Book and Dianping datasets. From Table~\ref{tab:inference layer}, we observe that recommendation performances increase first and then perform slightly drop when the graph inference layer increases. Specifically, \shortname~ achieves the best performance with $L=2$ on the Douban-Book dataset and $L=3$ on the Dianping dataset, respectively. This suggests that shallow graph inference layers can't well capture graph structure for node distribution estimation, but too deep graph inference layers also decrease the estimation quality due to the over-smoothing issue. 
\begin{table}[t]\centering
	\begin{small}
		\centering
		\setlength{\belowcaptionskip}{5pt}
		\caption{Performance on different graph inference layer $L$.}\label{tab:inference layer}
		\vspace{-8pt}
        \scalebox{1.1}{
		\begin{tabular}{|c|c|c|c|c|} \hline
		\multirow{2}{*}{Layers} &
		\multicolumn{2}{|c|}{Douban-Book}& \multicolumn{2}{|c|}{Dianping} \\
		\cline{2-5}
	&Recall@20&NDCG@20&Recall@20&NDCG@20\\ \hline
        L=1 & 0.1750 & 0.1555 & 0.1197 & 0.0733 \\ \hline
        L=2 & \textbf{0.1829} & \textbf{0.1638} & 0.1229 & 0.0751 \\ \hline
        L=3 & 0.1808 & 0.1618 & \textbf{0.1234} & \textbf{0.0757} \\ \hline
        L=4 & 0.1793 & 0.1605 & 0.1233 & 0.0752 \\ \hline
    	\end{tabular}}
	\end{small}
	\vspace{-0.2cm}
\end{table}

\textbf{Effect of Temperature $\tau$.}
As introduced in the previous works, temperature $\tau$ controls the mining scale of hard negatives~\cite{khosla2020supervised}. Specifically, a low temperature will highlight the gradient contributions of hard negatives that are similar to positive nodes. In \shortname~, there are two temperatures $\tau_1$ and $\tau_2$ in node-level and cluster-level contrastive losses, respectively. As suggested in the previous work, we fix the temperature $\tau_1=0.2$ on node-level contrastive loss, then analyze the impact of $\tau=\tau_2$ of the cluster-level contrastive loss. 
From Figure~\ref{fig: temp_and_prototype}(a) and Figure~\ref{fig: temp_and_prototype}(b), we have the following observations. First, too high or low temperature will decrease recommendation performance on all methods. A too-high temperature drops the ability to mine hard negative samples, while a too-low temperature will over-highlight hard negatives which are usually false negatives. 
Second, SGL and SimGCL achieve the best performances when temperature $\tau=0.2$ as suggested in the original paper, while \shortname~ achieves better performance on a smaller temperature, e.g., $\tau=0.13$ on Douban-Book and $\tau=0.15$ on Dianping dataset. The reason is that our proposed cluster-aware contrastive learning further encourages the consistency of nodes in a cluster, then a lower temperature will help the model better mine hard negatives.

 \begin{figure} [t]
  \begin{center}
      \subfigure[$\alpha$ on Douban-Book]{
      \includegraphics[width=40mm]{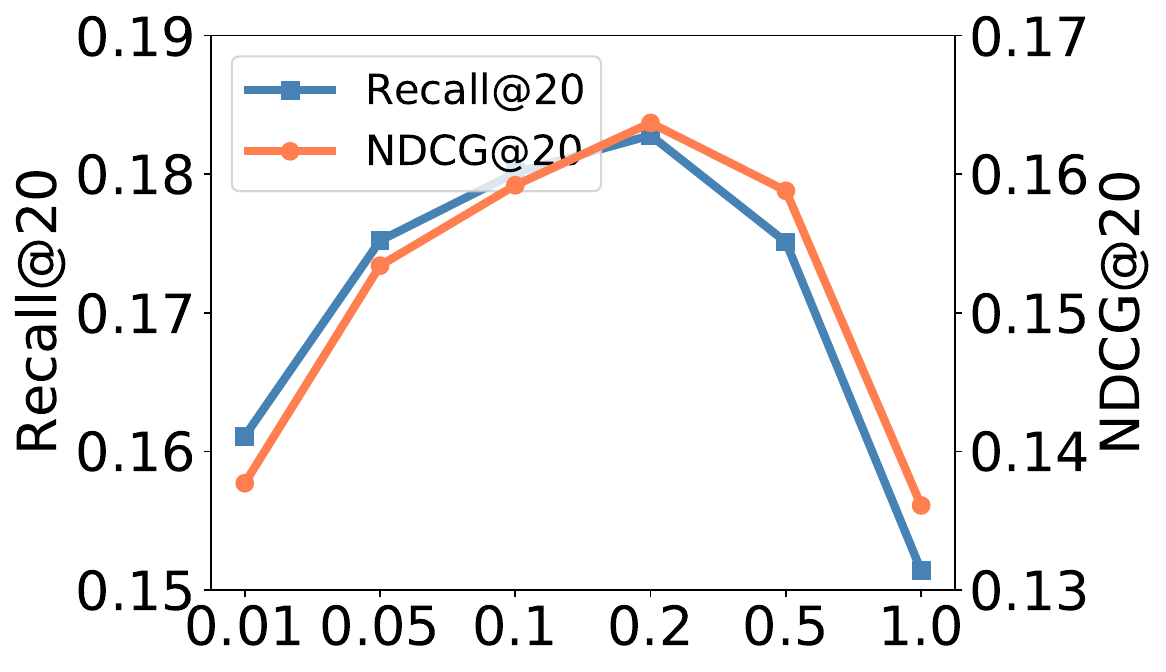}}
      \subfigure[$\gamma$ on Douban-Book]{
      \includegraphics[width=40mm]{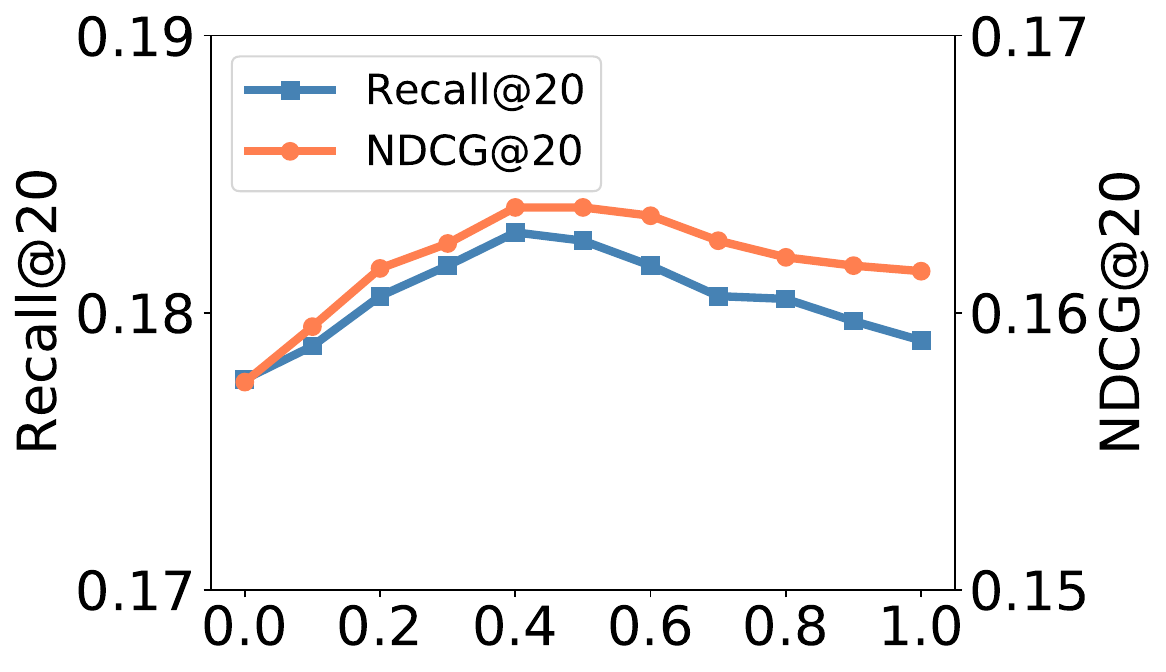}}
  \end{center}
  \vspace{-0.3cm}
  \caption{\small{Performance comparisons under different contrastie loss weights $\alpha$ and $\gamma$.}} 
  \label{fig: alpha}
  \vspace{-0.2cm}
\end{figure}

\textbf{Effect of Prototype Number $K_u$ and $K_i$.}
To investigate the effect of prototype numbers, we set the prototype numbers from zero to hundreds. We illustrate the experimental results in Figure\ref{fig: temp_and_prototype}(c) and Figure\ref{fig: temp_and_prototype}(d). Please note that when $K_u=K_i=0$, \shortname~ degenerates to VGCL-w/o C without the cluster-level objective. From this Figure, we find that \shortname~ consistently outperforms VGCL-C, which demonstrates that our proposed cluster-aware twofold contrastive learning strategy effectively improves the recommendation performance. For the Douban-Book dataset, \shortname~ reaches the best performances when $K_u=900$ and $K_i=300$. For the Dianping dataset, \shortname~ reaches the best performance when $K_u=500$ and $K_i=100$. It shows that precise clustering can provide pseudo-labels to distinguish contrastive samples.

\textbf{Effect of Contrastive Loss Weights $\alpha$ and $\gamma$.}
As illustrated in Figure\ref{fig: alpha}, we carefully tune the contrastive loss weights $\alpha$ and $\gamma$ on the Douban-Book dataset.
We observe that \shortname~ achieves the best performance when $\alpha=0.2$ and $\gamma=0.4$ on the Douban-Book dataset. As the space limit, we don't present analysis on the other two datasets, the best parameters are $\alpha=0.05, \gamma=0.5$ and $\alpha=0.1, \gamma=1.0$ on Dianping and Movielens-25M datasets, respectively.
Besides, the performance increases first and then drops quickly while $\alpha$ and $\gamma$ increase. It indicates that proper contrastive loss weights could effectively improve the sparse supervision issue, however, a too-strong self-supervised loss will lead to model optimization neglecting the recommendation task.

\section{Related Work}
\subsection{Graph based Collaborative Filtering}
Collaborative filtering is a popular technique widely used in recommender systems. The key is to learn user and item embeddings relying on historical interactions~\cite{hu2008collaborative,wu2022survey,shao2022faircf}. Early works leverage the matrix factorization technique to project users' and items' IDs into latent embeddings, and compute preferences with the inner product or neural networks~\cite{NIPS2008PMF, UAI2009BPR, WWW2017NCF}. Recently, borrowing the success of Graph Neural Networks~(GNNs)~\cite{ICLR2017Semigcn, NIPS2017inductive, NIPS2018GAT}, a series of graph-based models have been widely studied on various recommendation scenarios~\cite{ICLR2017GCMC, KDD2018PinSage, SIGIR2019NGCF, wu2020joint, wu2020learning, hu2021efficient}. As users' behavior can be naturally formulated as a user-item graph, graph-based CF methods formulate the high-order user-item graph structure on representation learning and achieve great performance improvements~\cite{SIGIR2019NGCF, AAAI2020LRGCCF, LightGCN, wang2021denoising, wang2021deconfounded}. NGCF is the first attempt that introduces GNNs to collaborative filtering, which injects high-order collaborative signals for embedding learning~\cite{SIGIR2019NGCF}. LR-GCCF proposes linear residual networks for user-item graph learning, which can effectively alleviate the over-smoothing issue in deep graph neural networks~\cite{AAAI2020LRGCCF}. LightGCN is a representative work and proposes a simplified graph convolution layer for CF which only has neighbor aggregation~\cite{LightGCN}. 
Despite effectiveness, graph-based CF methods also suffer from sparse supervision. In this work, we investigate collaborative filtering with self-supervised learning to tackle the above issues.

\subsection{Contrastive Learning based Recommendation}
As one of the popular self-supervised learning paradigms, contrastive learning aims to learn the representational invariants by data augmentation~\cite{jaiswal2020survey,liu2021self}. 
In general, contrastive learning first generates contrastive views from data augmentation, then maximize the mutual information to encourage the consistency of different contrastive views. Recently, some research successfully apply the CL technique to graph representation learning, either local-global scale contrast~\cite{ICLR2019DGI, ICLR2020infograph, shuai2022review} or global-global scale contrast~\cite{NIPS2020graph, you2021graph}. For instance, DGI learns node representations by maximizing the mutual information between the local and global representations~\cite{ICLR2019DGI}. GraphCL proposes four random graph augmentation strategies to multiple subgraphs for contrastive learning~\cite{NIPS2020graph}, AutoGCL further proposes an automated GCL method with learnable contrastive views~\cite{yin2022autogcl}. Inspired by these works, some GCL-based CF methods have been proposed~\cite{WSDM2021bipartite, yang2021egln, sgl, ncl, simgcl}.  BiGI maximizes the local-global mutual information on user-item bipartite graph~\cite{WSDM2021bipartite}. EGLN proposes to learn the enhanced graph structure and maximize the mutual information maximization with a local-global objective~\cite{yang2021egln}. Besides, data augmentation based CL techniques are usually applied to CF, aiming to deal with sparse supervision and noise interaction problems~\cite{sgl,ncl,simgcl}. SGL designs three structure graph augmentations to generate contrastive views and improve recommendation accuracy and robustness by maximizing the consistency of different views~\cite{sgl}. NCL proposes neighborhood-enriched contrastive learning to improve performance, it uses the correlated structure neighbors and semantic neighbors as contrastive objects~\cite{ncl}. SimGCL revisits structure augmentation methods and proposes a simple feature augmentation to enhance GCL-based recommendations~\cite{simgcl}. 

Despite the effectiveness, we argue that current GCL-based recommendation methods are still limited by data augmentation strategies whatever structure or feature augmentation. First, structure augmentation randomly deletes nodes or edges of the input graph to generate subgraphs for contrastive learning. However, random structure augmentation is easy to destroy the intrinsic nature of the original user-item graph. Besides, feature augmentation adds the same scale noise to all nodes, ignoring the unique characteristics of nodes~(such as degree on the graph), thus can't satisfy all nodes. In this work, we propose a novel contrastive paradigm without data augmentation and implement adaptive contrastive loss learning for different nodes.



\subsection{VAE and Applications on Recommendation}
Variational Auto-Encoder~(VAE) is a generative method widely used in machine learning~\cite{kingma2013vae,rezende2014stochastic}. It assumes that the input data can be generated from variables with some probability distribution. Following, some extensions of VAE are proposed to improve performance from different perspectives~\cite{sohn2015learning,higgins2016beta,im2017denoising,ivanov2018variational}. CVAE considers complex condition distribution on inference and generation process~\cite{sohn2015learning}, $\beta$-VAE proposes to learn the disentangled representations by adding the loss of KL-term~\cite{higgins2016beta}, and DVAE reconstructs the input data from its corrupted version to enhance the robustness~\cite{im2017denoising}. The basic idea of applying VAEs to the recommendation is to reconstruct the input users' interactions. Mult-VAE proposes that multinomial distribution is suitable for modeling user-item interactions, and parameterizes users by neural networks to enhance the representation ability~\cite{Multi-VAE}. RecVAE further improves Mult-VAE by introducing a novel composite prior distribution for the latent encoder~\cite{shenbin2020recvae}. BiVAE proposes bilateral inference models to estimate the user-item distribution and item-user distribution~\cite{truong2021bilateral}. CVGA combines GNNs and VAE and proposes a novel collaborative graph auto-encoder recommendation method, which reconstructs user-item bipartite graph using variance inference~\cite{zhang2022revisiting}.
Besides, some works attempt to leverage VAEs for sequential recommendation ~\cite{xie2021adversarial} and cross-domain recommendation~\cite{salah2021towards}. Different from the above VAE-based recommendation models, our \shortname~ introduces the variational inference technique to generate multiple contrastive views for GCL-based recommendation, which build a bridge between generative and contrastive learning models for recommendation.

\section{Conclusion}
In this work, we investigate GCL-based recommendation from the perspective of better contrastive view construction, and propose a novel \fullname~ framework. Instead of data augmentation, we leverage the variational graph reconstruction technique to generate contrastive views to serve contrastive learning. Specifically, we first estimate each node's probability distribution by graph variational inference, then generate contrastive views with multiple samplings from the estimated distribution. As such, we build a bridge between the generative and contrastive learning models for recommendation. The advantages have twofold. First, the generated contrastive representations can well reconstruct the original graph without information distortion. Second, the estimated variances vary from different nodes, which can adaptively regulate the scale of contrastive loss for each node. Furthermore, considering the similarity of the estimated distributions of nodes, we propose a cluster-aware twofold contrastive learning, a node-level to encourage consistency of a node's contrastive views and a cluster-level to encourage consistency of nodes in a cluster. Empirical studies on three public datasets clearly show the effectiveness of the proposed framework.

\section*{Acknowledgements}
This work was supported in part by grants from the National Key Research and Development Program of China (Grant No. 2021ZD0111802), the National Natural Science Foundation of China (Grant No. 72188101, 61932009, 61972125, U19A2079, 62006066, U22A2094), Major Project of Anhui Province (Grant No. 202203a05020011), and the CCF-AFSG Research Fund (Grant No. CCF-AFSG RF20210006).

\balance
\bibliographystyle{ACM-Reference-Format}
\bibliography{VGCL}
\appendix
\end{document}